\documentclass[10pt, a4paper]{article}
\usepackage{lrec}
\usepackage{multibib}
\newcites{languageresource}{Language Resources}
\usepackage{graphicx}
\usepackage{tabularx}
\usepackage{soul}

\usepackage{epstopdf}
\usepackage[utf8]{inputenc}

\usepackage{hyperref}
\usepackage{xstring}

\usepackage{color}

\usepackage{amsmath,graphicx}
\usepackage{amssymb,amsmath,epsfig}
\usepackage{lipsum,color,multirow,url}
\usepackage{arydshln, algorithm, algpseudocode}
\usepackage{microtype}
\usepackage{nth}
\usepackage[inline]{enumitem}
\usepackage{romannum}

\title{Multilingual Graphemic Hybrid ASR with Massive Data Augmentation}

\name{Chunxi Liu,  Qiaochu Zhang, Xiaohui Zhang, Kritika Singh,    Yatharth Saraf, Geoffrey Zweig}

\address{Facebook AI \\
         New York, NY, and Menlo Park, CA, USA \\
         \{chunxiliu,frankz,xiaohuizhang,skritika,ysaraf,gzweig\}@fb.com  }

\abstract{
Towards developing high-performing ASR for low-resource languages, approaches to address the lack of resources are to make use of data from multiple languages, and to augment the training data by creating acoustic variations. 
In this work we present a single grapheme-based ASR model learned on 7 geographically proximal languages, using standard hybrid BLSTM-HMM acoustic models with lattice-free MMI objective. 
We build the single ASR grapheme set via taking the union over each language-specific grapheme set, and 
we find such multilingual graphemic hybrid ASR model can perform language-independent recognition on all 7 languages, and substantially outperform each monolingual ASR model.
Secondly, we evaluate the efficacy of multiple data augmentation alternatives within language, as well as their complementarity with multilingual modeling. 
Overall, we show that the proposed multilingual graphemic hybrid ASR with various data augmentation can not only recognize any within training set languages, but also provide large ASR performance improvements.  \\ 
\newline  }

\begin{document}

\maketitleabstract

\section{Introduction}
\label{sec:intro}

It can be challenging to build high-accuracy automatic speech recognition (ASR) systems in the real world due to the vast language diversity and the requirement of extensive manual annotations on which the ASR algorithms are typically built. 
Series of research efforts have thus far been focused on guiding the ASR of a target language by using the supervised data from multiple languages.  

Consider the standard hidden Markov models (HMM) based hybrid ASR system with a phonemic lexicon, where the vocabulary is specified by a pronunciation lexicon. One popular strategy is to make all languages share the same phonemic representations through a universal phonetic alphabet such as International Phonetic Alphabet (IPA) phone set \cite{lin2009study,liu2016adapting,pulugundla2018but,tong2019investigation}, or X-SAMPA phone set \cite{wells1995computer,knill2013investigation,knill2014language,wiesner2018automatic}. 
In this case, multilingual joint training can be directly applied.
Given the effective neural network based acoustic modeling,  another line of research is to share the hidden layers across multiple languages while the softmax layers are language dependent \cite{huang2013cross,heigold2013multilingual}; 
such multitask learning procedure can improve ASR accuracies for both within training set languages, and also unseen languages after language-specific adaptation, i.e.,  cross-lingual transfer learning. 
Different nodes in hidden layers have been shown in response to distinct phonetic features \cite{nagamine2015exploring}, and hidden layers can be potentially transferable across languages.  
Note that the above works all assume the test language identity to be known at decoding time, and the language specific lexicon and language model applied. 

In the absence of a phonetic lexicon, building graphemic systems has shown comparable performance to phonetic lexicon-based approaches in extensive monolingual evaluations \cite{kanthak2002context,gales2015unicode,trmal2017kaldi}. 
Recent advances in end-to-end or sequence-to-sequence ASR models have attempted to take the union of multiple language-specific grapheme (i.e. orthographic character) sets, and use such union as a universal grapheme set for a single sequence-to-sequence ASR model \cite{watanabe2017language,toshniwal2018multilingual,kim2018towards,kannan2019large}. 
It allows for learning a grapheme-based model jointly on data from multiple languages, and performing ASR on within training set languages. 
In various cases it can produce performance gains over monolingual modeling that uses in-language data only. 

Since HMM-based hybrid model remains a competitive ASR approach especially in low/medium-resource settings \cite{luscher2019rwth,wang2019transformer},
in our work, we aim to examine the same approach above of building a multilingual graphemic lexicon, while using a hybrid ASR system -- based on Bidirectional Long Short-Term Memory (BLSTM) and HMM -- learned with lattice-free maximum mutual information (MMI) objective \cite{povey2016purely}.
Our initial attempt is on building a single cascade of an acoustic model, a phonetic decision tree, a graphemic lexicon and a language model
-- for 7 geographically proximal languages that have little overlap in their character sets.
We evaluate it in a low resource context where each language has around 160 hours training data. 
We find that, despite the lack of explicit language identification (ID) guidance, our multilingual graphemic hybrid ASR model can accurately produce ASR transcripts in the correct test language scripts, and provide higher ASR accuracies than each language-specific ASR model. 
We further examine if using a subset of closely related languages -- along language family or orthography -- can achieve the same performance improvements as using all 7 languages.

Though extensive end-to-end or  sequence-to-sequence ASR works have been built on multilingual graphemic models, to the best of our knowledge, there is no prior work in hybrid ASR that uses a single multilingual graphemic lexicon (rather than an IPA or X-SAMPA based phonetic lexicon) for multiple training languages. 
In this work, we show for the first time that  multilingual graphemic hybrid ASR can provide  large improvements across all training languages, even though almost each training language has distinct graphemic set.

We proceed with our investigation on various data augmentation techniques to overcome the lack of training data in the above low-resource setting.  
Given the highly scalable neural network acoustic modeling, 
extensive alternatives to increasing the amount or diversity of existing training data have been explored in prior works, e.g., applying vocal tract length perturbation and speed perturbation  \cite{ko2015audio}, volume perturbation and normalization \cite{peddinti2015jhu},  additive noises \cite{amodei2016deep},  reverberation \cite{peddinti2015jhu,ko2017study,kim2017generation},    and  SpecAugment  \cite{park2019specaugment}. 
In this work we focus particularly on techniques that mostly apply to our  wildly collected  video datasets. 
In comparing their individual and complementary effects, we aim to answer:
(i) if there is benefit in scaling the model training to significantly larger quantities, e.g., up to 9 times greater than the original training set size, and 
(ii) if any, is the data augmentation efficacy comparable or complementary with the above multilingual modeling.

Improving accessibility to videos ``in the wild” such as automatic captioning on YouTube has been studied in  \cite{liao2013large,soltau2016neural}. 
While allowing for applications like video captions, indexing and retrieval,  
transcribing the heterogeneous social media videos of extensively diverse languages is highly challenging for ASR systems.
On the whole, we present empirical studies in building a single multilingual graphemic hybrid ASR model capable of language-independent decoding on multiple languages, and in effective data augmentation techniques for video datasets.

\section{Multilingual  Graphemic Hybrid ASR}
\label{sec:asr}

In this section we first briefly describe our deployed hybrid ASR architecture based on the weighted finite-state transducers (WFSTs) outlined in \cite{mohri2008speech}.
Then we present its extension to multilingual training. 
Lastly, we discuss its language-independent decoding and language-specific decoding. 

\subsection{Graphemic ASR with WFST}

In the ASR framework of a hybrid BLSTM-HMM,   the decoding graph can be interpreted as a composed WFST of cascade $H \circ C \circ L \circ G$.  
Acoustic models, i.e. BLSTMs, produce acoustic scores over context-dependent HMM (i.e. triphone) states. A WFST $H$, which represents the HMM set, maps the triphone states to context-dependent phones. 

While in graphemic ASR, the notion of phone is turned to grapheme, and we typically create the grapheme set via modeling each orthographic character as a separate grapheme. 
Then a WFST $C$ maps each context-dependent grapheme, i.e. tri-grapheme, to an orthographic character. 
The lexicon $L$ is specified where each word is mapped to a sequence of characters forming that word.
$G$ encodes either the transcript during training, or a language model during decoding.

\subsection{A Single Multilingual ASR Model Using Lattice-Free MMI}
\label{subsec:multi_asr}

To build a single grapheme-based acoustic model for multiple languages, 
a multilingual graphemic set is obtained by taking a union of each grapheme set from each language considered, each of which can be either overlapping or non-overlapping. 
In the multilingual graphemic lexicon, each word in any language is mapped to a sequence of characters in that language.

A context-dependent acoustic model is constructed using the decision tree clustering of tri-grapheme states, in the same fashion as the context dependent triphone state tying \cite{young1994tree}. 
The graphemic-context decision tree is constructed over all the multilingual acoustic data including each language of interest.  
The optimal number of leaves for the multilingual model tends to be larger than for a monolingual neural network.

The acoustic model is a BLSTM network, using sequence discriminative training with lattice-free MMI objective \cite{povey2016purely}.
The BLSTM model is bootstrapped from a standard Gaussian mixture model (GMM)-HMM system. 
A multilingual $n$-gram language model is learned over the combined transcripts including each language considered.

\subsection{Language-Independent and  Language-Specific Decoding in the WFST Framework}
\label{subsec:specific_decode}

Given the multilingual lexicon and language model, the multilingual ASR above can decode any within training set language, even though not explicitly given any information about language identity. We refer to it as language-independent decoding or multilingual decoding. Note that such ASR can thus far  produce any word in the multilingual lexicon, and the hypothesized word can either be in the vocabulary of the considered test language, or out of test language vocabulary as a mismatched-language error.

We further consider applying  language-specific  decoding, assuming the test language identity to be known at decoding time. 
Again consider the decoding graph  $H \circ C \circ L \circ G$,  and $H$ \& $C$ are thus multilingual while the lexicon $L$ and language model  $G$ can  include only the words in test language vocabulary. 
The multilingual acoustic model can therefore make use of multilingual training data, while its  language-specific decoding operation only produces monolingual words matched with test language identity.    


\section{Data Augmentation}
\label{sec:data_augmentation}

In this section, we consider 3 categories of data augmentation techniques that are effectively applicable to video datasets.

\subsection{Speed and Volume Perturbation}
\label{subsec:speed}

Both speed and volume perturbation emulate mean shifts in spectrum \cite{ko2015audio,peddinti2015jhu}.
To perform speed perturbation of the training data, we produce three versions of each audio with speed factors $0.9$, $1.0$, and $1.1$.  The training data size is thus tripled. 
For volume perturbation, each audio is scaled with a random variable drawn from a uniform distribution $[0.125, 2]$.

%

\subsection{Additive Noise}
\label{subsec:additive}

To further increase training data size and diversity,  we can create new audios via superimposing each original audio with additional noisy audios in time domain. 
To obtain diverse noisy audios, we use AudioSet, which consists of 632 audio event classes and a collection of over 2 million manually-annotated 10-second sound clips from YouTube videos \cite{gemmeke2017audio}. 

Note that in our video datasets, video lengths vary between 10 seconds and 5 minutes, with an average duration of about 2 minutes. 
Rather than constantly repeating the 10-second sound clip to match the original minute-long audio, we superimpose each sound clip on the short utterances via audio segmentation.
Specifically,  we first use an initial bootstrap model to align each original long audio, and segment each audio into around 10-second utterances via word boundaries. 

Then for each utterance in the original train set,  we can create a new noisy utterance by the steps:
\begin{enumerate}         
\item   Sample a sound clip from AudioSet. 
\item   Trim or repeat the sound clip as necessary to match the duration of the original utterance.
\item   Sample a signal-to-noise ratio (SNR) from a Gaussian distribution with mean 10, and round the SNR up to 0 or down to 20 if the sample is beyond 0-20dB. Then scale the sound clip signal to obtain the target SNR.  
\item   Superimpose the original utterance signal with the scaled sound clip signal in time domain to create the resulting utterance. 
\end{enumerate}
\noindent 
Thus for each original utterance, we can create a variable number of new noisy utterances via sampling sound clips. We use a 3-fold augmentation that combines the original train set with two noisy copies.

\subsection{SpecAugment}
\label{subsec:spec}

We consider applying the frequency and time masking techniques -- which are shown to greatly improve the performance of end-to-end ASR models \cite{park2019specaugment} --  to our hybrid systems. 
Similarly, they can be applied online during each epoch of LF-MMI training, while time warping requires the need for realignment and thus does not fit hybrid model training.  

Consider each utterance (i.e. after the audio segmentation in Section \ref{subsec:additive}), and we compute its log mel spectrogram with $\nu$ dimension   and  $\tau$  time steps:  
\begin{enumerate}   
\item Frequency masking is applied $m_F$  times, and each time the frequency bands $\lbrack f_0$, $f_0+ f)$ are masked, where $f$ is sampled from $\lbrack 0, F\rbrack$ and  $f_0$ is sampled from $\lbrack 0, \nu - f)$.     
\item Time masking is optionally applied $m_T$ times, and each time the time steps $\lbrack t_0$, $t_0+ t)$ are masked,  where  $t$ is sampled from $\lbrack 0, T\rbrack$ and   $t_0$   is sampled from $\lbrack 0, \tau - t)$.   
\end{enumerate}
\noindent 
As in \cite{park2019specaugment}, we increase the training schedule accordingly, i.e., number of epochs. 

\setlength{\tabcolsep}{0.1cm}
\begin{table}[t]
\caption{\label{tab:data} {\it  The amounts of audio data in hours.}}
\centering 
\begin{tabular}{   c  c  c c c c  }
\hline \hline
        Language               &       Train      &    \multicolumn{4}{   c  }{Test }  \\ 
          &     &   \texttt{clean}          &       \texttt{noisy}    &     \texttt{xtrm\Romannum{1}}    &    \texttt{xtrm\Romannum{2}}   \\
\hline \hline
\multirow{1}{*}{ }    Kannada         & 125.5       &       1.5      &    9.9         &  0.8         &  2.7    \\
 			    Malayalam       & 127.7      &       4.5      &     9.2        &     0.7       &    1.0  \\    
                             Sinhala            & 160.0         &     13.9     &     25.0     &     8.6        &  8.8  \\  
                             Tamil               & 176.9       &       2.8       &    16.4        &       0.5     &   0.7   \\  
 \hline \hline
                            Bengali             &   160.0     &    7.4         &     24.9     &    25.0      &  16.4             \\  
                            Hindi                 &   160.0     &   22.2         &      21.5    &    19.4      & 19.8     \\ 
                           Marathi            &   148.6       &      2.7         &    13.7    &    0.3       & 0.5      \\  
\hline \hline
\end{tabular}
\end{table}
\setlength{\tabcolsep}{0.132cm}
\begin{table*} [!htb]  
\caption{\label{tab:results} 
{\it  WER results on each video dataset. Frequency masking is denoted by fm, speed perturbation by sp, and additive noise (Section \ref{subsec:additive}) by noise.   
3lang, 4lang and 7lang denote the multilingual ASR models trained on 3, 4 and 7 languages, respectively, as in Section \ref{subsection:data}. 
\textit{Lang-specific decoding} denotes using multilingual acoustic model with language-specific lexicon and language model,  as in Section \ref{subsec:specific_decode}.    
Average is unweighted average WER across 4 video types. 
Gain (\%) is the relative reduction in the Average WER over each monolingual baseline. 
}
}
\centerline{ 
\begin{tabular}{  p{1.5cm}  p{6.2cm}  |  p{1.1cm}  p{1.1cm} p{1.1cm} p{1.1cm}   | p{1.0cm}   | p{1.1cm}  }
\hline \hline
                  Language            &   Model                 &   \texttt{clean}       &    \texttt{noisy}   &     \texttt{xtrm\Romannum{1}}    &      \texttt{xtrm\Romannum{2}}  &  Average  & \%  Gain  \\ 
\hline \hline
 \multirow{1}{*}{ Kannada }   &   monolingual                            							   &    56.9       &   56.6             &   58.7  	&   57.6   &   57.5      & \ \  --   \\
                                             &    monolingual  \texttt{+}  fm      							   &    53.3      &    54.8          &   56.9  		&   56.4     &  55.4     &   3.7   \\      
                                             &    monolingual  \texttt{+}  sp       							   &     53.1     &     54.7         &   56.4     	&  55.2     &  54.9   &     4.5      \\     
                                             &    monolingual  \texttt{+}  fm   \texttt{+}   sp      			    	  & 50.3         &    53.1         &    \textbf{54.8}      &    53.9    &    53.0    &     7.8   \\     
                                             &    monolingual  \texttt{+}  sp   \texttt{+}   noise     			  &   50.7         &   53.3              &  54.8     	&    53.6    &   53.1   &          7.7    \\ 
                                             &    monolingual  \texttt{+}  fm   \texttt{+}   sp   \texttt{+}  noise   &   \textbf{49.7}     &    \textbf{52.5}    &    54.9 &   \textbf{52.7}     &  \textbf{52.5}  &       \textbf{8.7}   \\      \cline{2-8}   
                                             &    multilingual, \emph{4lang}         							&    	50.2                &   \textbf{53.4}       &   	55.7             &   \textbf{53.4}       &    \textbf{53.2}   &    \textbf{7.5}   \\
                                             &    multilingual, \emph{7lang}        								 &    	\textbf{49.7}    &    53.5                  &  \textbf{54.9}        &    55.6                &   53.4                 &    7.1             \\           \cdashline{2-8}[1.0pt/0.5pt]  
                                             &    multilingual, \emph{7lang}         	  \texttt{+}   \textit{lang-specific decoding}		 &   \textit{49.4}   &    \textit{52.5}     &  	\textit{54.6}   &     \textit{53.7}    &   \textit{52.5}    &     \textit{8.7}       \\   
                                             &    multilingual, \emph{7lang}     \texttt{+}   fm   \texttt{+}   sp   \texttt{+}  noise              &    	\textbf{46.6}    &  \textbf{52.0}      &   \textbf{53.0}     &    \textbf{53.3}     &  \textbf{51.2}       &     \textbf{11.0}   \\
\hline
\multirow{1}{*}{ Malayalam }   &   monolingual                             &   56.5      &   53.2     &  70.3            &   55.9   &  59.0        &   \ \   --  \\ 
					     &    multilingual, \emph{4lang}           		 &   	52.8     &   \textbf{51.6}    &   \textbf{65.8}           &   \textbf{53.4}     &  \textbf{55.9}   &   \textbf{5.3}   \\
 					     &    multilingual, \emph{7lang}           		  &    \textbf{52.1}    &    51.9   &   	66.3      &   54.0     &  56.1   &     5.0    \\
\hline
\multirow{1}{*}{ Sinhala }    &   monolingual                               &      45.4      &     39.5        &   62.7     &    	51.8   &    49.9         &  \ \   --    \\ 
					 &    multilingual, \emph{4lang}            		&    	\textbf{42.1}    &    	38.4   &  	59.7      &    	50.3    &   47.6      &  4.6     \\
					 &    multilingual, \emph{7lang}            		&     	42.9   &   \textbf{38.3}   &   \textbf{59.3}     &   	\textbf{49.9}     &  \textbf{47.6}  &    \textbf{4.6}  \\
\hline
\multirow{1}{*}{ Tamil }        &   monolingual                               &    44.2       &         44.4             &    49.0           &   52.7        &    47.6         &   \ \   --     \\ 
					 &    multilingual, \emph{4lang}            		&    	40.7    &  	42.8       &   	46.6            &     \textbf{50.9}       &      45.2   &       5.0        \\
					 &    multilingual, \emph{7lang}            		&   	\textbf{40.1}     &    	\textbf{42.7}     &   	\textbf{46.1}             &    51.7       &       \textbf{45.2}   &   \textbf{5.0}     \\
\hline \hline  
\multirow{1}{*}{ Bengali }   &   monolingual                                  &    53.4     &     50.8         &   68.2            &  58.0       &  57.6   &     \ \   --    \\ 
					&    multilingual, \emph{3lang}          	        &      	\textbf{45.5}    &     \textbf{47.0}        &    \textbf{62.6}            &  \textbf{53.3}      &   \textbf{52.1}  &     \textbf{9.5}   \\
					 &    multilingual, \emph{7lang}           		 &    	45.7    &     48.1        &   63.9          &    54.7      &   53.1  &      7.8   \\
\hline
\multirow{1}{*}{ Hindi }       &   monolingual                             							  &  36.9       &    38.2    &    58.4    &    45.0       &    44.6           &       \ \   --        \\ 
					&    monolingual    \texttt{+}  fm      							&     33.2     &    34.8       &   54.1     &    40.9    & 40.8          &    8.5   \\
					&    monolingual    \texttt{+}  sp      							 &      33.6    &     34.9      &   55.0    &    41.1    &   41.2         &    7.6   \\
					&    monolingual    \texttt{+}  fm   \texttt{+}   sp    		 		   &    32.1      &    33.4       &  52.7   &    39.5    &  39.4          &    11.7     \\
					&    monolingual    \texttt{+}  sp   \texttt{+}   noise     				  &    32.0      &     33.5      &  52.6   &     39.5   &    39.4       &    11.7       \\
					&    monolingual    \texttt{+}   fm   \texttt{+}   sp   \texttt{+}  noise        &     \textbf{30.9}    &     \textbf{32.2}      &   \textbf{50.7}     &     \textbf{38.2}   & \textbf{38.0}        &   \textbf{14.8}     \\     \cline{2-8}  
					&    multilingual, \emph{3lang}           	    							 &     	32.2                         &     	33.9                    &   	\textbf{53.5}      &     	\textbf{40.3}   &  40.0                             &    10.3     \\
					 &    multilingual, \emph{7lang}          	   							  &    \textbf{31.9}            &      \textbf{33.8}            &        53.6             &   40.8                &  \textbf{40.0}         &   \textbf{10.3}    \\           \cdashline{2-8}[1.0pt/0.5pt]  
					 &    multilingual, \emph{7lang}        \texttt{+}    \textit{lang-specific decoding}             &   \textit{31.8}     &    \textit{33.4}    &  \textit{52.7}     &      \textit{40.1}     &   \textit{39.5}                     &    \textit{11.4}   \\      
					 &    multilingual, \emph{7lang}     \texttt{+}   fm   \texttt{+}   sp   \texttt{+}  noise       &   \textbf{28.5}     &   \textbf{30.8}      &     \textbf{49.6}    &      \textbf{36.7}      &      \textbf{36.4}           &    \textbf{18.4}     \\      
\hline
\multirow{1}{*}{ Marathi }   &   monolingual                               &   38.2     &     39.8           &   	63.2       &   49.0   &    47.6                                                                      &      \ \   --     \\ 
					&    multilingual, \emph{3lang}           	     &    \textbf{34.9}      &   	\textbf{37.4}           &    \textbf{56.4}     &   46.3     &    \textbf{43.7}         &     \textbf{8.2}   \\
					 &    multilingual, \emph{7lang}                    &    35.2      &      38.1          &      56.5      &     \textbf{46.1}   &  44.0                                            &    7.6    \\
\hline \hline
\end{tabular}}
\end{table*}

\section{Experiments}
\label{sec:exp}
\vspace{-0.0001cm}

\subsection{Data}
\label{subsection:data}

Our multilingual ASR attempt was on 7 geographically proximal languages: Kannada, Malayalam, Sinhala, Tamil,  Bengali, Hindi and Marathi. 
The datasets were a set of public social media videos, which were wildly collected and anonymized.  We categorized them into four sets:   
\texttt{clean}, \texttt{noisy}, 
\texttt{extreme\Romannum{1}} (\texttt{xtrm\Romannum{1}}) and 
\texttt{extreme\Romannum{2}}   (\texttt{xtrm\Romannum{2}}). 
\texttt{xtrm\Romannum{1}}     differed from       \texttt{xtrm\Romannum{2}}   in chronological order, and were both more acoustically challenging than \texttt{clean} and \texttt{noisy} categories.

For each language, the train and test set size are described in Table \ref{tab:data}, and most training data were of \texttt{noisy} category. 
On each language we also had a small validation set for model parameter tuning. 
Each monolingual ASR baseline was trained on language-specific data only. 

To create the grapheme set,  we consult the unicode character ranges of each language, and also include apostrophe, hyphen and zero width joiner in the final character sets. 
The character sets of these 7 languages have little overlap except that (i) they all include common basic Latin alphabet, and (ii) both Hindi and Marathi use Devanagari script.
We took the union of 7 character sets therein as the multilingual grapheme set (Section \ref{subsec:multi_asr}), which contained 432 characters.       
In addition, we deliberately split 7 languages into two groups, such that the languages within each group were more closely related 
in terms of language family, orthography or phonology.
We thus built 3 multilingual ASR models trained on: 
\begin{enumerate} [label=(\roman*)]      
\item   all 7 languages, for 1059 training hours in total, 
\item   4 languages -- Kannada, Malayalam, Sinhala and Tamil -- for  590 training hours, 
\item   3 languages -- Bengali, Hindi and Marathi --  for  469 training hours, 
\end{enumerate}
\noindent  which are referred to as \emph{7lang}, \emph{4lang}, and \emph{3lang} respectively.  
Note that Kannada, Malayalam and Tamil  are  Dravidian languages, which have rich agglutinative inflectional morphology \cite{pulugundla2018but} and resulted in around 10\% OOV token rates on test sets 
(Hindi had the lowest OOV rate as 2-3\%). 
Such experimental setup was designed to answer the questions: 
\begin{enumerate} [label=(\roman*)]  
\item  If a single graphemic ASR model could scale its language-independent recognition up to  all 7 languages.   
\item  If including all 7 languages could yield better ASR performance than using a small subset of closely related languages. 
\end{enumerate}
\subsection{Model Configurations}
\label{subsection:model}

Each bootstrap model was a GMM-HMM based system with speaker adaptive training, implemented with Kaldi \cite{povey2011kaldi}.
Each neural network acoustic model was a latency-controlled BLSTM \cite{zhang2016highway}, learned with lattice-free MMI objective and Adam optimizer \cite{kingma2014adam}.   
All neural networks were implemented with Caffe2 \cite{hazelwood2018applied}. 
Due to the production real time factor (RTF) requirements, we used the same model size in all cases -- a 4 layer latency-controlled BLSTM network with 600 cells in each layer and direction -- except that, 
the softmax dimensions, i.e. the optimal decision tree leaves, were determined through experiments on validation sets, varying within 7-30k. 
Input acoustic features were 80-dimensional log-mel filterbank coefficients. 
After lattice-free MMI training, the model with the best accuracy on validation set was used for evaluation on test set.
We used standard 5-gram language models in all cases.  Each multilingual 5-gram language model is learned simply via combining transcripts of each language.

\subsection{Results with Multilingual ASR}

ASR word error rate (WER\%) results are shown in Table \ref{tab:results}. 
We found that, although not explicitly given any information on test language identities,  multilingual ASR with language-independent decoding  (Section  \ref{subsec:specific_decode}) - trained on 3, 4, or 7 languages - substantially outperformed each monolingual ASR in all cases,  and on average led to relative WER reductions between 4.6\% (Sinhala) and 10.3\% (Hindi). 

Note that, in contrast to the multilingual phonetic hybrid ASR (i.e. using phonetic lexicons),  it is intuitive to see ASR performance improve when different languages share the same phone set via IPA or X-SAMPA ,
since each phonetic modeling  can use more training data than monolingual training.
However, in our multilingual graphemic ASR, only 2 of 7 training languages overlapped in  character sets;  
for the first time, we show that,   such multilingual graphemic-context decision tree  based hybrid ASR can still improve performance for all languages.


Also, the word hypotheses from language-independent decoding could be language mismatched, e.g., part of a Kannada utterance was decoded into Marathi words. 
So we counted how many word tokens in the decoding transcripts were not in the lexicon of corresponding test language. We found in general only 1-3\% word tokens are language mismatched, indicating that the multilingual model was very effective in identifying the language implicitly and jointly recognizing the speech.

Consider the scenario that,  test language identities are known likewise in each monolingual ASR, and we proceed with language-specific decoding (Section  \ref{subsec:specific_decode}) on Kannada and Hindi, via language-specific lexicon and language model at decoding time. We found that, the language-specific decoding  provided only moderate gains, presumably as  discussed above,  the language-independent decoding had given the mismatched-language word token rates as sufficiently low as 1-3\%.

Additionally, the multilingual ASR of \emph{4lang}  and \emph{3lang}  (Section \ref{subsection:data}) achieved the same, or even slightly better performance as compared to the ASR of \emph{7lang}, 
suggesting that incorporating closely related languages into multilingual training is most useful for improving ASR performance. 
However, the \emph{7lang}  ASR  by itself still yields the advantage in language-independent recognition of more languages. 
\subsection{Results with Data Augmentation}

First, we experimented with monolingual ASR on Kannada and Hindi, 
and performed comprehensive evaluations of the data augmentation techniques described in Section \ref{sec:data_augmentation}.
As in Table \ref{tab:results}, the performance gains of using frequency masking were substantial and comparable to those of using speed perturbation, where $m_F = 2$ and $F=15$ (Section \ref{subsec:spec}) worked best.       
In addition, combining both frequency masking and speed perturbation could provide further improvements.   
However, applying  additional volume perturbation (Section \ref{subsec:speed})  or  time masking (Section \ref{subsec:spec})  was not helpful in our monolingual experiments, and we omit showing the results in the table. 


Note that after speed perturbation, the training data tripled, to which we could apply another 3-fold augmentation based on additive noise (Section \ref{subsec:additive}), 
and the final train set was thus 9 times the size of original train set.  
We found that all 3 techniques were complementary, and in combination led to large fusion gains over each monolingual baseline  --  relative WER reductions of  8.7\% on Kannada, and 14.8\%  on Hindi.   

%
%

Secondly, we applied the 3 data augmentation techniques to the multilingual ASR of \emph{7lang}, and tested their additive effects.    
We show the resulting WERs on Kannada and Hindi in Table \ref{tab:results}. 
Note that on Kannada,  we found around 7\% OOV token rate on  \texttt{clean}    but around 10-11\% on other 3 test sets,   and we observed more gains on  \texttt{clean} ; 
presumably because the improved acoustic model could only correct the in-vocabulary word errors,   lower OOV rates therefore left more room for improvements. 
Hindi had around 2.5\% OOV rates on each test set, and we found incorporating data augmentation into multilingual ASR led to on average  
9.0\% relative WER reductions. 


Overall, we demonstrated the multilingual hybrid ASR with massive data augmentation -- via a single graphemic model even without the use of explicit language ID -- allowed for  relative WER reductions of 
11.0\% on Kannada and  18.4\% on Hindi. 

\section{Conclusion}

Multilingual training have been extensively studied in conventional phonetic hybrid ASR  \cite{lin2009study,knill2013investigation} and the recent end-to-end ASR \cite{watanabe2017language,toshniwal2018multilingual}. 
In our work, for the first time, we demonstrate that a multilingual grapheme-based hybrid ASR model can effectively perform language-independent recognition on any within training set languages, and substantially outperform each monolingual ASR alternative. 
Various data augmentation techniques can yield further complementary improvements. 
Such single multilingual model can not only provide better ASR performance, but also serves as an alternative to a typical production deployment, which typically includes extensive monolingual ASR systems and a separate language ID model. 
The proposed approach of building a single multilingual graphemic hybrid ASR model without requiring  individual language ID 
- while being  especially competitive in low-resource settings  - can greatly simplify the productionizing  and maintenance process.

Additionally, as compared to the multilingual multitask learning plus monolingual fine-tuning methods in \cite{huang2013cross,heigold2013multilingual}, 
our preliminary experimentation shows that our proposed approach above can give comparable performance without requiring  separate language ID guidance during decoding. We leave the detailed studies to the future work. 
Also, future work will  expand the  language coverage to include both geographically proximal and distant languages.



\section{Acknowledgements}

The authors would like to thank Duc Le, Ching-Feng Yeh and  Siddharth Shah, all with Facebook, for their  invaluable infrastructure assistance  and technical discussions.     We also thank  Yifei Ding  and Daniel McKinnon, also at Facebook, for coordinating the ASR language expansion efforts.


\section{Bibliographical References}\label{reference}

\bibliographystyle{lrec}
\bibliography{lrec2020W-xample-kc,thesis,refs}

\label{lr:ref}
\bibliographystylelanguageresource{lrec}
\bibliographylanguageresource{languageresource}

\end{document}